# ACO-OFDM-Specified Recoverable Upper Clipping With Efficient Detection for Optical Wireless Communications

Wei Xu, *Member, IEEE*, Man Wu, Hua Zhang, *Member, IEEE*,
Xiaohu You, *Fellow, IEEE*, and Chunming Zhao, *Member, IEEE*

National Mobile Communications Research Laboratory, Southeast University,
Nanjing 210096, China

DOI: 10.1109/JPHOT.2014.2352643
1943-0655 © 2014 IEEE. Translations and content mining are permitted for academic research only.
Personal use is also permitted, but republication/redistribution requires IEEE permission.
See http://www.ieee.org/publications_standards/publications/rights/index.html for more information.

Manuscript received July 25, 2014; revised August 13, 2014; accepted August 14, 2014. Date of publication August 28, 2014; date of current version September 10, 2014. This work was supported in part by the 973 Program under Grant 2013CB329204 and in part by the National Natural Science Foundation of China (NSFC) under Grant 61471114 and Grant 61223001. Corresponding author: W. Xu (e-mail: wxu@seu.edu.cn).

**Abstract:** The high peak-to-average-power ratio (PAPR) of orthogonal frequency-division multiplexing (OFDM) degrades the performance in optical wireless communication systems. This paper proposes a modified asymmetrically clipped optical OFDM (ACO-OFDM) with low PAPR via introducing a recoverable upper-clipping (RoC) procedure. Although some information is clipped by a predetermined peak threshold, the clipped error information is kept and repositioned in our proposed scheme, which is named RoC-ACO-OFDM, instead of simply being dropped in conventional schemes. The proposed method makes full use of the specific structure of ACO-OFDM signals in the time domain, where half of the positions are forced to zeros within an OFDM symbol. The zero-valued positions are utilized to carry the clipped error information. Moreover, we accordingly present an optimal maximum *a posteriori* (MAP) detection for the RoC-ACO-OFDM system. To facilitate the usage of RoC-ACO-OFDM in practical applications, an efficient detection method is further developed with near-optimal performance. Simulation results show that the proposed RoC-ACO-OFDM achieves a significant PAPR reduction, while maintaining a competitive bit-error rate performance compared with the conventional schemes.

## 1. Introduction

Optical wireless communications (OWC), including its indoor and outdoor applications, have attracted an increasing interest in the field of next generation wireless transmission technologies [1], [2]. To meet the ultra-high data rate challenge of emerging multimedia services, OWC enjoys both a unique advantage of extraordinarily large amount of unlicensed optical spectrum and its immunity to the ever-growing radio frequency (RF) coverage. These beneficial attributes promise OWC as a powerful alternative to RF systems in future wireless networks, especially for "last-mile" access. Potential applications of OWC have been shown ubiquitous for various scenarios, e.g., ultra-dense networks, small cell coverage, secrecy communications, internet-of-things, etc.

A nature of optical communications is that the information is modulated in the optical domain which restricts transmit signals to be nonnegative values. Meanwhile, it is difficult to collect appreciable signal power in a single electromagnetic mode at an optical receiver end. Therefore, intensity modulation and direct detection (IM/DD) [3]–[5] becomes a common way of optical modulation, as exemplified by using on-off keying (OOK) in [5]. Although OOK is simple for implementation, its power efficiency is low since it only carries a single binary information via turning the lights on and off. Orthogonal frequency division multiplexing (OFDM) has been acknowledged as an effective modulation technique for high data rate communication systems including asymmetric digital subscriber lines (ADSL) and new generation cellular networks, namely long term evolution (LTE). Recently, OFDM has also been applied to optical communication systems due to its advantages of high power efficiency and ability of alleviating inter-symbol interference (ISI) [6]–[10].

In order to guarantee nonnegative signals for optical modulation, there are generally two types of OFDM techniques tailored for OWC, namely direct-current-biased optical OFDM (DCO-OFDM) [8] and asymmetrically-clipped optical OFDM (ACO-OFDM) [9]. By applying the Hermitian symmetry of the Fourier transform, the optical OFDM outputs can be guaranteed as real values. In order to further make the signals nonnegative, DCO-OFDM adds a DC bias to the optical OFDM outputs before transmission, while ACO-OFDM clips the negative outputs at zeros and transmits the remaining positive parts. In [10], comprehensive comparisons between DCO-OFDM and ACO-OFDM have been conducted from both theoretical and numerical perspectives. Although either technique has its own beneficial and detrimental features, both DCO- and ACO-OFDM face the same challenge of high peak-to-average power ratio (PAPR), which is recognized as a main defect for all OFDM systems [11]. In optical OFDM systems, however, the issue of high PAPR becomes more crucial because it leads to a nonlinear distortion not only caused by power amplifiers but also by a light emitting diode (LED) emitter [12]. To achieve reduced PAPR, a number of approaches have been reported, as summarized in the overviews [11] and [13], for OFDM in RF applications. Compared to the relatively mature research in RF OFDM, the PAPR reduction technique has recently been studied for optical systems [14]–[16], [18]–[20]. Particularly in [14], a block coding based scheme was proposed by extending its usage from RF scenarios to optical ones. In [15] and [16], a modification of the selective mapping, which was referred to as SLM in [17] for RF OFDM, was presented to fit optical applications. By inserting some dedicated pilots for transmission pattern recognition, the authors in [18] developed an adaptive transmission strategy to reduce clipping loss. The pilot-assisted method achieves better PAPR performance than the SLM method at high-level constellations, while it suffers effective data rate loss especially under dense pilot patterns. To the best of our knowledge, most of the existing techniques, however, are inherited from their RF counterparts with necessary modifications to fit optical channels.

Focusing on the optical OFDM, source data mapping to subcarriers is adjusted for attaining the Hermitian symmetry to generate real OFDM signals. Due to this difference from an ordinary mapping in RF OFDM, the PAPR of optical OFDM showed its unique characterizations compared with the PAPR behaviors in RF applications [21]. By applying the features of the specifically-structured ACO/DCO-OFDM, the previous work [19] formulated a quadratic form of the PAPR specified for DCO-OFDM and accordingly developed a semi-definite program (SDP) based PAPR optimization with an iterative mechanism. Although the DC bias of DCO-OFDM could cause a more strict requirement on PAPR, the problem of PAPR reduction is also challenging for ACO-OFDM. Consider the nonlinearity of LED, as exemplified in [22], it is observed that the dynamic range of a linear LED could be within $0.3 < x < 0.8$. This implies that, even for ACO-OFDM, a bias is better to be added to avoid the nonlinearity for small signals. Moreover, putting aside the LED nonlinearity nature, the ACO-OFDM generally suffers a more severe PAPR problem than DCO-OFDM because the designed structure of ACO-OFDM. Therefore, in this paper, we propose a modified ACO-OFDM with low PAPR via introducing a recoverable upper-clipping (RoC) procedure. Even though some information is clipped according to the allowed peak values, the clipped error information is kept and repositioned in our proposed scheme, named RoC-ACO-OFDM, instead of simply being dropped in conventional schemes. The proposed method makes full use of the specific structure of ACO-OFDM signals in time domain where a half of the positions are forced to zeros within an OFDM symbol. The zero-valued positions are utilized to carry the clipped error information, which helps the receiver to recover the clipped signal before detection. Notice that this RoC procedure inevitably complicates the detection of ACO-OFDM. We subsequently presented an optimal maximum *a posteriori* (MAP) detection for the RoC-ACO-OFDM system. Moreover, to further facilitate the usage of our proposed RoC-ACO-OFDM in practical applications, an efficient detection method is also developed with near optimal performance. Simulation results finally verify the efficiency and effectiveness of the proposed scheme with both optimal and simplified detection methods.

The remainder of this paper is structured as follows. In Section 2, a description of conventional ACO-OFDM and its properties are introduced. In Section 3, our proposed RoC-ACO-OFDM is presented with an optimal MAP detection. Subsequently in Section 4, a low complexity detection method is further developed with efficient computations. Numerical results are illustrated in Section 5 before concluding remarks in Section 6.

## 2. Optical Wireless Communications Using ACO-OFDM

### 2.1. System Model of ACO-OFDM

As one of the major ways of implementing optical OFDM, ACO-OFDM enjoys a high power efficiency at a cost of efficiency in frequency (subcarrier) domain [9]. In fact, only a quarter of all subcarriers in an ACO-OFDM are utilized to carry information. More specifically, assume that the input information was modulated as a group of $N/4$ complex symbols $\{S(I), I = 0, 1, \ldots, N/4 - 1\}$ using a chosen modulation like *M*-quadrature amplitude modulation (QAM), phase-shift keying (PSK), etc. Mapping the complex source symbols to an *N*-subcarrier ACO-OFDM follows two rules in order to ensure nonnegative OFDM outputs for optical modulation. First, Hermitian symmetry is imposed on subcarriers to ensure that the OFDM signals in time domain are real values after inverse fast Fourier transform (IFFT). Secondly, the modulated symbols are only mapped onto odd subcarriers while all even subcarriers are set to zeros. This second rule generates OFDM samples of the first half time domain indices copied in the second half part with their signs flipped.

Fig. 1 illustrates the diagram of an ACO-OFDM symbol in both frequency and time domains. After mapping data to subcarriers shown in Fig. 1(a), data at the *k*th subcarrier of an ACO-OFDM symbol follows

$$X(k) = \begin{cases} S\left(\frac{k-1}{2}\right), & k \text{ is odd, and } k < \frac{N}{2} \\ S^*\left(\frac{N-k-1}{2}\right), & k \text{ is odd, and } k > \frac{N}{2} \\ 0, & \text{otherwise} \end{cases} \quad (1)$$

where the upperscript $(\cdot)^*$ represents conjugate of a complex value. The mapping with Hermitian symmetry ensures that output of IFFT, saying $x[n] = \text{IFFT}(X(k))$, is real time-domain signals. Moreover, since all even subcarriers are set to be zeros, an important property of anti-symmetry of $x[n]$ is further achieved [23]. It gives

$$x[n] = -x[n + N/2], \quad n = 0, 1, \ldots, \frac{N}{2} - 1 \quad (2)$$

which indicates that the pair of time-domain samples $(x[n], x[n + N/2])$ shares a same value with opposite signs, as shown in Fig. 1(b). Note that this is crucial for ACO-OFDM since no information is actually lost during the procedure when clipping all negative parts at zeros. The finally obtained ACO-OFDM signal after clipping the negative samples at zeros is given by

$$x_c[n] = \begin{cases} x[n], & x[n] \geq 0 \\ 0, & \text{otherwise} \end{cases} \quad (3)$$

which is exemplified in Fig. 1(c). In this way, any pair $(x[n], x[n+N/2])$ of the time-domain samples of an ACO-OFDM symbol has one nonnegative value remained while the other one is forced to zero. After adding cyclic prefix (CP) and passing through a digital-to-analog converter (DAC), the symbol is emitted using a LED via optical modulation.

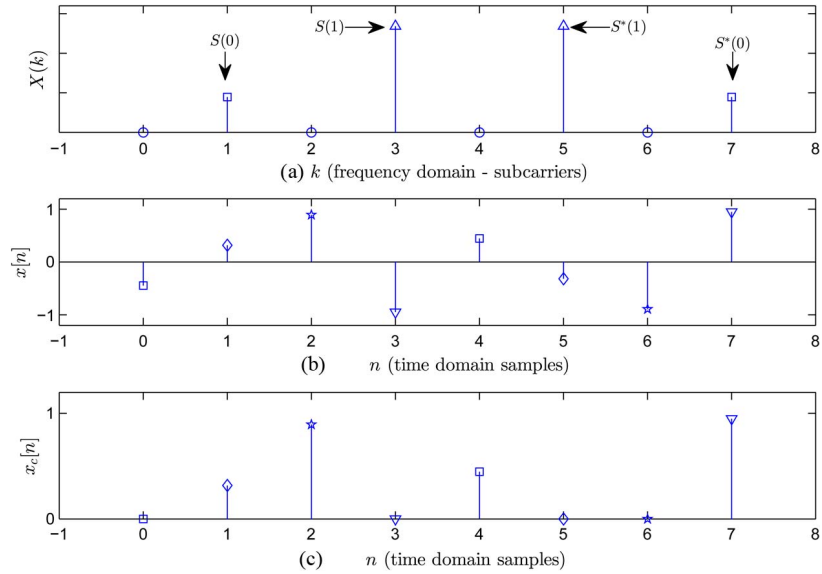

Fig. 1. Signal samples in both frequency and time domains of an ACO-OFDM.

At the receiver side, the received signals can be generally modeled as

$$y_c[n] = x_c[n] + z[n] \quad (4)$$

where $z[n]$ is white Gaussian noise with zero mean and variance of $\sigma_z^2$. Conventional detect and demodulation methods for OFDM can be directly applied for ACO-OFDM [9] with subtle adjustments. According to the transmitter design of ACO-OFDM, a half of the received samples should definitely be zeros. In [24], this feature was taken into account for enhancing the ACO-OFDM detector by forcing negative received samples to zeros, which reports a 1.25 dB gain at very high signal-to-noise ratios (SNRs). In [23], this feature was considered with detection in a more sophisticated way. A pairwise maximum likelihood (ML) detector was proposed in [23] and only the detected clipped position in a pair, i.e., $(\tilde{y}_c[n], \tilde{y}_c[n+N/2])$, was forced to zero for demodulation. Moreover, there are other detectors proposed for ACO-OFDM, like in [25] and [26], achieving similar performance. Considering both complexity and performance, we choose the ML-based approach [23] in our simulation parts for comparison.

### 2.2. Statistical Properties of ACO-OFDM Signals

This subsection reviews some statistical results on ACO-OFDM, which will be shown useful in the following analysis of our proposed RoC-ACO-OFDM. By applying the central limit theorem, the behavior of $x[n]$ can be approximated by Gaussian distribution because $x[n]$ can be treated as a sum of a large number of independent random variables. This is accurate for a large OFDM size. Note that this Gaussian approximation is accurate for large $N \geq 64$ and that it can be easily verified by numerical examples. The assumption has been reported and widely used, e.g., in [6], [11], [13], [17], [23], and [25], for the sake of analytical simplicity with guaranteed accuracy. Hence, it is directly to know that the clipped signal $x_c[n]$ in (3) follows the half Gaussian with its probability density function (PDF) given by [23]

$$f_{x_c}(x) = 0.5\delta(x) + \frac{u(x)}{\sqrt{2\pi}\sigma_x} e^{\frac{-x^2}{2\sigma_x^2}} \tag{5}$$

where $\delta(\cdot)$ is the Dirac delta function, $u(\cdot)$ is the Heaviside step function, and $\sigma_x$ is the standard deviation satisfying

$$\mathbb{E}\{|x_c[n]|^2\} = \int_{-\infty}^{\infty} x^2 f_{x_c}(x) dx = \frac{\sigma_x^2}{2} \tag{6}$$

where $\mathbb{E}\{\cdot\}$ denotes the expectation operator. Denote $\hat{x}_c[n] = x_c[n + N/2]$ for notational simplicity. In the following parts, we treat the pair of ACO-OFDM samples $(x_c[n], \hat{x}_c[n])$ as a basic element for data processing. From [23], the joint PDF of the sample pair $(x_c[n], \hat{x}_c[n])$ follows:

$$f_{x_c, \hat{x}_c}(x_1, x_2) = \frac{u(x_1)}{\sqrt{2\pi}\sigma_x} e^{\frac{-x_1^2}{2\sigma_x^2}} \delta(x_2) + \frac{u(x_2)}{\sqrt{2\pi}\sigma_x} e^{\frac{-x_2^2}{2\sigma_x^2}} \delta(x_1). \tag{7}$$

## 3. Recoverable Upper-Clipping With Optimal Detection

### 3.1. Proposed ACO-OFDM Transmission With RoC Procedure

In a conventional ACO-OFDM, the samples $\{x_c[n], n = 0, 1, \ldots, N-1\}$ are directly modulated on optical waves for transmission. As is well-known in OFDM, the direct transmission faces a challenge of high PAPR, especially for the ACO-OFDM systems [12]. Noticeable performance degradation occurs due to the effect of high PAPR. Clipping is a popular way of PAPR reduction for both RF and optical OFDM systems. However, inevitable information loss is caused during the clipping procedure, which may also degrade the system performance. Therefore, tradeoffs between clipping and PAPR reduction are always attractive focuses in OFDM design [12]–[15]. A number of PAPR reduction approaches for RF OFDM have recently been tailored for optical systems, while few has well explored the specific structures of optical OFDM. In this section, we present a modified ACO-OFDM with an additional upper-clipping procedure. By properly exploiting the specific time domain structure of ACO-OFDM signals, the proposed upper-clipping procedure can fortunately allow the clipped information being well recovered at receiver side while in existing schemes the clipping information is always dropped without possible recovery.

In the proposed RoC-ACO-OFDM, we introduce an additional upper clipping procedure at the transmitter as well as a corresponding signal recovery at the receiver. As shown in Fig. 2, a block diagram of the proposed RoC-ACO-OFDM is depicted with additional parts highlighted in a dashed box. Now we are ready to elaborate the details of the proposed upper-clipping in RoC-ACO-OFDM. Let $\eta_c$ denote an upper-clipping level which implies that the upper-clipping is triggered once the sample value exceeds $\eta_c$, or equivalently, let $\text{CR} = 20\log(\eta_c/\sqrt{\mathbb{E}\{|x_c[n]|^2\}})$ be the common alternative clipping ratio (CR) representing the clipping level in dB [6]. The upper-clipping of RoC is designed for clipping the ACO-OFDM signals on a pairwise basis, that is the clipping is carried out for each sample pair $(x_c[n], \hat{x}_c[n])$ with $n = 0, 1, \ldots, N/2 - 1$.

Take a given sample pair $(x_c[n], \hat{x}_c[n])$ for instance. Without loss of generality, we assume that $x_c[n]$ carries the nonnegative information data and hence $\hat{x}_c[n]$ is a zero according to the

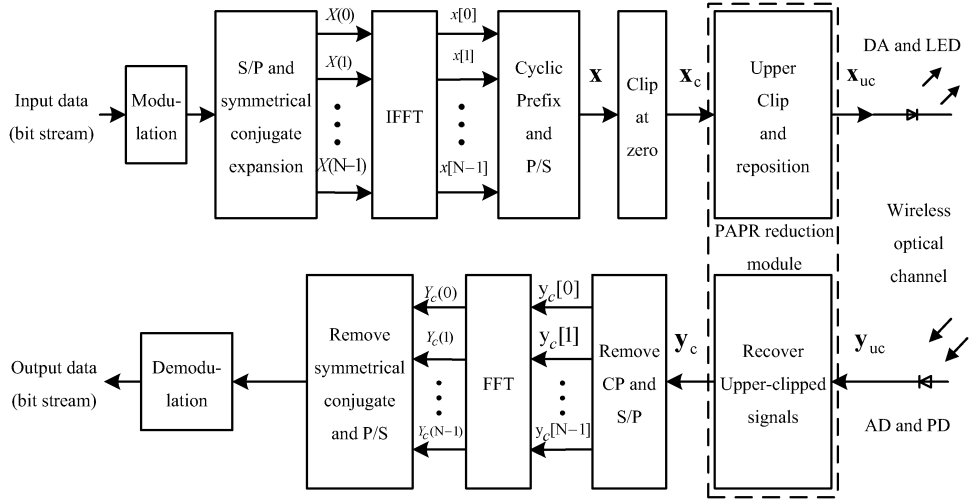

Fig. 2. Block diagram of the transmitter and receiver for RoC-ACO-OFDM. S/P: serial to parallel; P/S: parallel to serial; DA: digital to analog; AD: analog to digital.

properties in (2) and (3). The upper-clipping at $\eta_c$ is trivial, as follows:

$$x_c[n] = \begin{cases} \eta_c, & x_c[n] > \eta_c \\ x_c[n], & \text{otherwise} \end{cases} \tag{8}$$

which yields $e[n] = x_c[n] - \eta_c > 0$ referred to as the clipped error information under the case of $x_c[n] > \eta_c$. Instead of simply being dropped, it is interesting to find that the clipped error information $e[n]$ can be kept and repositioned at the other place of the sample pair if a proper recovery detection can be designed at the receiver of ACO-OFDM. Because the other sample $\hat{x}_c[n] = 0$ bears no information, the proposed RoC procedure replaces the zero $\hat{x}_c[n]$ with $e[n]$. Thus, applying RoC on $(x_c[n], \hat{x}_c[n])$ yields a new sample pair $(x_{uc}[n], \hat{x}_{uc}[n])$ given by

$$(x_{uc}[n], \hat{x}_{uc}[n]) = \begin{cases} (\eta_c, e[n]), & x_c[n] > \eta_c \\ (x_c[n], 0), & \text{otherwise.} \end{cases} \tag{9}$$

A manipulation of the RoC procedure is exemplified in Fig. 3. In this way, no information is lost during the RoC procedure and the obtained vector $\mathbf{x}_{uc} = [x_{uc}[0], x_{uc}[1], \ldots, x_{uc}[N-1]]$ is then ready to be transmitted after optical modulation, as shown in Fig. 2.

After transmission over optical channels, the corresponding received sample pair can be expressed by

$$(y_{uc}[n], \hat{y}_{uc}[n]) = (x_{uc}[n], \hat{x}_{uc}[n]) + (z[n], \hat{z}[n]), \quad n = 0, 1, \ldots, N-1 \tag{10}$$

where we define $\hat{y}_{uc}[n] = y_{uc}[n + N/2]$. For conventional ACO-OFDM systems, the received signal can be effectively detected and demodulated by exploiting the *a priori* information that either sample in the received pair $(y_{uc}[n], \hat{y}_{uc}[n])$ should be zero if correctly detected. In the RoC-ACO-OFDM, however, this *a priori* information never holds after applying the RoC procedure at the transmitter. Two sample values in the pair can be both positive. Therefore, in order to ensure signal recovery before OFDM demodulation, the receiver needs to detect which position in a pair $(y_{uc}[n], \hat{y}_{uc}[n])$ is the corresponding one being upper clipped at the transmitter. To facilitate this detection before signal recovery, we introduce another threshold $\alpha\eta_c$ in the RoC procedure where $\alpha \in (0, 1)$ is a predetermined fixed coefficient. Note that for a sample pair in (9) there is still a probability that $e[n] > \eta_c$. The occurrence of $e[n] > \eta_c$ not only violates the clipping at $\eta_c$ but also makes it harder for receiver to detect the clipped position in a sample pair. We introduce the threshold $\alpha\eta_c$ to further clip $e[n]$ before being repositioned. The purpose of this process is to ensure that the repositioned $e[n]$, e.g., in (9), is no larger than $\alpha\eta_c$. As will be shown in the following subsection, this process does help the receiver detect the clipped position in a received pair before signal recovery.

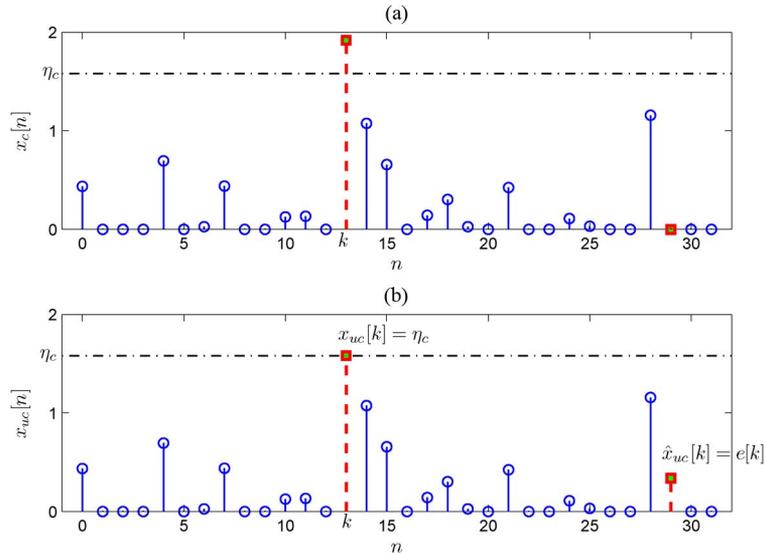

Fig. 3. Example of manipulation of RoC procedure from $x_c[n]$ to $x_{uc}[n]$.

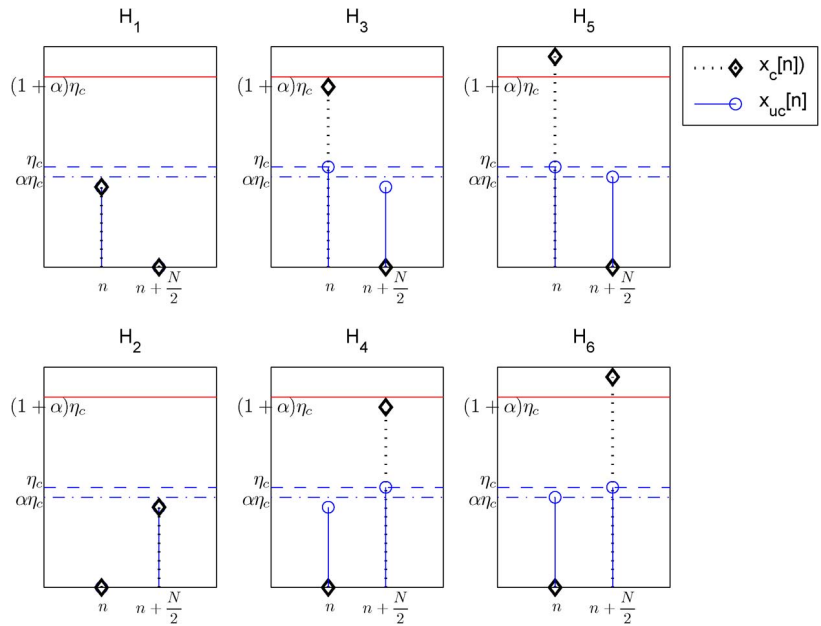

Fig. 4. Six upper-clipping cases of the proposed RoC in ACO-OFDM systems.

Based on the above descriptions, we can accordingly divide the RoC procedure on $(x_c[n], \hat{x}_c[n])$ into six different cases, as illustrated in Fig. 4. From this figure, $H_1$ and $H_2$ represent two cases where no sample is larger than $\eta_c$ in the pair $(x_c[n], \hat{x}_c[n])$, and hence it directly yields $(x_{uc}[n], \hat{x}_{uc}[n]) = (x_c[n], \hat{x}_c[n])$ where no sample is upper-clipped. $H_3$ and $H_4$ exemplify the cases where a sample in the pair is larger than $\eta_c$ and it is clipped to $\eta_c$. While for very special cases where a sample is even larger than $(1+\alpha)\eta_c$, or equivalently, $e[n] > \alpha\eta_c$, the clipping procedures are illustrated as $H_5$ and $H_6$ in Fig. 4. In summary, the proposed RoC procedure can be mathematically expressed as

$$(x_{uc}[n], \hat{x}_{uc}[n]) = \begin{cases} (x_c[n], 0), & 0 < x_c[n] \leq \eta_c \\ (0, \hat{x}_c[n]), & 0 < \hat{x}_c[n] \leq \eta_c \\ (\eta_c, e[n]), & \eta_c < x_c[n] \leq (1+\alpha)\eta_c \\ (\hat{e}[n], \eta_c), & \eta_c < \hat{x}_c[n] \leq (1+\alpha)\eta_c \\ (\eta_c, \alpha\eta_c), & x_c[n] > (1+\alpha)\eta_c \\ (\alpha\eta_c, \eta_c), & \hat{x}_c[n] > (1+\alpha)\eta_c \end{cases} \quad (11)$$

where $\hat{e}[n] = \hat{x}_c[n] - \eta_c$ denotes the clipped error information on $\hat{x}[n]$, and the last two cases in (11), corresponding to $H_5$ and $H_6$ in Fig. 4, imply that the clipped error information $e[n]$ or $\hat{e}[n]$ is large enough to be clipped to $\alpha\eta_c$ before being repositioned.

### 3.2. MAP Detection With Recovery for RoC-ACO-OFDM

In conventional ACO-OFDM systems, the receiver first detects whether the transmit symbol is of the types $(x_c[n], 0)$ or $(0, \hat{x}_c[n])$ based on the received $(y_c[n], \hat{y}_c[n])$ corrupted by noise. After completing detections for all received pairs, a traditional OFDM demodulation is applied by using FFT. From Fig. 2, the transmit symbol $\mathbf{x}_{uc}$ of the proposed RoC-ACO-OFDM is an upper-clipped version of $\mathbf{x}_c$ as expressed in (11). Hence, given a received pair $(y_{uc}[n], \hat{y}_{uc}[n])$, the corresponding transmit pair $(x_{uc}[n], \hat{x}_{uc}[n])$ can be any one of the six possible cases in (11), making the detection more complicated than that for a conventional ACO-OFDM, which is obviously is a crucial challenge of the proposed RoC-ACO-OFDM. In the following, we present an MAP based detection algorithm with signal recovery for RoC-ACO-OFDM.

Before calculating the probabilities for MAP detection, we first define the following hypotheses indicating different types of transmitted sample pairs as exemplified in Fig. 4. From (11), six hypotheses of $(x_{uc}[n], \hat{x}_{uc}[n])$ are defined as follows:

$$\begin{aligned} &H_1: \ (x_{uc}[n], \hat{x}_{uc}[n]) = (x, 0); &&H_2: \ (x_{uc}[n], \hat{x}_{uc}[n]) = (0, x) \\ &H_3: \ (x_{uc}[n], \hat{x}_{uc}[n]) = (\eta_c, x - \eta_c); &&H_4: \ (x_{uc}[n], \hat{x}_{uc}[n]) = (x - \eta_c, \eta_c) \\ &H_5: \ (x_{uc}[n], \hat{x}_{uc}[n]) = (\eta_c, \alpha\eta_c); &&H_6: \ (x_{uc}[n], \hat{x}_{uc}[n]) = (\alpha\eta_c, \eta_c) \end{aligned} \quad (12)$$

where $x \geq 0$ is an arbitrary sample value at the transmitter. Without loss of generality, we consider the received sample pair $(y_{uc}[n], \hat{y}_{uc}[n])$ for a fixed $n$. Given that the received values of the pair equals $(y_1, y_2)$, the MAP detection of the transmission type of $(x_{uc}[n], \hat{x}_{uc}[n])$ can be formulated as follows:

$$\hat{H} = \arg_{H_i} \max_{H_i, i=1,2,\ldots,6} P_{H|(y_{uc}[n], \hat{y}_{uc}[n])}\{H_i|(y_1, y_2)\} \quad (13)$$

where $P_{H|(y_{uc}[n], \hat{y}_{uc}[n])}$ is the probability of the occurrence $H$ conditioned on the received sample pair $(y_{uc}[n], \hat{y}_{uc}[n])$. In order to evaluate the detection in (13), we apply the Bayes theorem and obtain the following equality:

$$P_{H|(y_{uc}[n], \hat{y}_{uc}[n])}\{H_i|(y_1, y_2)\} = \frac{P_H\{H_i\} \times f_{y_{uc}[n], \hat{y}_{uc}[n]|H}(y_1, y_2|H_i)}{f_{y_{uc}[n], \hat{y}_{uc}[n]}(y_1, y_2)} \quad (14)$$

where $f_{y_{uc}[n], \hat{y}_{uc}[n]}(\cdot)$ and $f_{y_{uc}[n], \hat{y}_{uc}[n]|H}(\cdot)$ are respectively the joint PDF and the joint PDF conditioned on $H$ of $(y_{uc}[n], \hat{y}_{uc}[n])$. Because the probability density $f_{y_{uc}[n], \hat{y}_{uc}[n]}(y_1, y_2)$ is the same for all hypothesis $H_i$, the MAP detection in (13) by substituting (14) equals

$$\hat{H} = \arg_{H_i} \max_{H_i, i=1,2,\ldots,6} g(H_i, y_1, y_2) \quad (15)$$

where we define the function $g(H_i, y_1, y_2) = P_H\{H_i\} \times f_{y_{uc}[n], \hat{y}_{uc}[n]|H}(y_1, y_2|H_i)$ for notational simplicity.

To calculate the function of $g(H_i, y_1, y_2)$, we first consider the probability $P_H\{H_i\}$ of six hypotheses. Let us denote two constants as follows:

$$p_1 = \int_{\eta_c}^{(1+\alpha)\eta_c} \frac{1}{\sqrt{2\pi}\sigma_x} e^{\frac{-x^2}{2\sigma_x^2}} dx, \quad p_2 = \int_{(1+\alpha)\eta_c}^{\infty} \frac{1}{\sqrt{2\pi}\sigma_x} e^{\frac{-x^2}{2\sigma_x^2}} dx. \tag{16}$$

Consider the probability of $H_1$ in (12) which equals the probability of $0 \leq x \leq \eta_c$. From the PDF of $x$ given in (5), we have

$$P_H(H_1) = \frac{1}{2} - p_1 - p_2. \tag{17}$$

Due to the symmetry of $H_1$ and $H_2$, it directly gives $P_H(H_2) = P_H(H_1)$. Similarly for other hypotheses, it is easy to get

$$P_H(H_3) = P_H(H_4) = p_1, \quad P_H(H_5) = P_H(H_6) = p_2. \tag{18}$$

Regarding the conditional joint PDF $f_{y_{uc}[n],\hat{y}_{uc}[n]|H}(y_1, y_2 | H_i)$ and from the system model in (10), it can be calculated as

$$f_{y_{uc},\hat{y}_{uc}|H}(y_1, y_2 | H_i) = \int_{-\infty}^{\infty}\int_{-\infty}^{\infty} f_{x_{uc},\hat{x}_{uc}|H}(y_1 - z_1, y_2 - z_2 | H_i) f_{z,\hat{z}}(z_1, z_2) \, dz_1 dz_2 \tag{19}$$

where

$$f_{z,\hat{z}}(z_1, z_2) = \frac{1}{2\pi\sigma_z^2} e^{-\frac{z_1^2 + z_2^2}{2\sigma_z^2}} \tag{20}$$

is the joint PDF of two independent Gaussian variables, and $f_{x_{uc},\hat{x}_{uc}|H}(\cdot,\cdot)$ is the joint PDF of a transmit sample pair under a given hypothesis $H$. Now according to the hypotheses defined in (12) and from the joint PDF in (7), the conditional joint PDFs can be expressed by

$$f_{x_{uc},\hat{x}_{uc}|H_1}(x_1, x_2) = f_{x_{uc},\hat{x}_{uc}|H_2}(x_1, x_2) = \frac{u(x_1) - u(x_1 - \eta_c)}{\sqrt{2\pi}\sigma_x P_H(H_1)} e^{-\frac{x_1^2}{2\sigma_x^2}} \delta(x_2)$$

$$f_{x_{uc},\hat{x}_{uc}|H_3}(x_1, x_2) = f_{x_{uc},\hat{x}_{uc}|H_4}(x_1, x_2) = \frac{u(x_2) - u(x_2 - \alpha\eta_c)}{\sqrt{2\pi}\sigma_x P_H(H_3)} e^{-\frac{(x_2+\eta_c)^2}{2\sigma_x^2}} \delta(x_1 - \eta_c)$$

$$f_{x_{uc},\hat{x}_{uc}|H_5}(x_1, x_2) = f_{x_{uc},\hat{x}_{uc}|H_6}(x_1, x_2) = \delta(x_1 - \eta_c)\delta(x_2 - \alpha\eta_c) \tag{21}$$

where the equalities also use the symmetry property of the hypotheses $H_1$ and $H_2$, $H_3$ and $H_4$, and $H_5$ and $H_6$. By substituting (20) and (21) into (19) and after some integral manipulations, the conditioned joint PDF of $f_{y_{uc},\hat{y}_{uc}|H}(\cdot,\cdot)$ can be rewritten as

$$f_{y_{uc},\hat{y}_{uc}|H_1}(y_1, y_2) = \frac{1}{2\pi\sigma_z\sqrt{\sigma_x^2+\sigma_z^2}P_H(H_1)} \exp\left(-\frac{y_1^2\sigma_z^2 + y_2^2(\sigma_x^2+\sigma_z^2)}{2\sigma_z^2(\sigma_x^2+\sigma_z^2)}\right)\left[Q\left(\frac{y_1 - \eta_c - \frac{y_1\sigma_z^2}{\sigma_x^2+\sigma_z^2}}{\sigma}\right) - Q\left(\frac{y_1 - \frac{y_1\sigma_z^2}{\sigma_x^2+\sigma_z^2}}{\sigma}\right)\right]$$

$$f_{y_{uc},\hat{y}_{uc}|H_3}(y_1, y_2) = \frac{1}{2\pi\sigma_z\sqrt{\sigma_x^2+\sigma_z^2}P_H(H_3)} \exp\left(-\frac{(y_2+\eta_c)^2\sigma_z^2 + (y_1-\eta_c)^2(\sigma_x^2+\sigma_z^2)}{2\sigma_z^2(\sigma_x^2+\sigma_z^2)}\right)$$
$$\times \left[Q\left(\frac{y_2 - \alpha\eta_c - \frac{(y_2+\eta_c)\sigma_z^2}{\sigma_x^2+\sigma_z^2}}{\sigma}\right) - Q\left(\frac{y_2 - \frac{(y_2+\eta_c)\sigma_z^2}{\sigma_x^2+\sigma_z^2}}{\sigma}\right)\right]$$

$$f_{y_{uc},\hat{y}_{uc}|H_5}(y_1, y_2) = \frac{1}{2\pi\sigma_z^2} \exp\left(-\frac{(y_1-\eta_c)^2 + (y_2-\alpha\eta_c)^2}{2\sigma_z^2}\right) \tag{22}$$

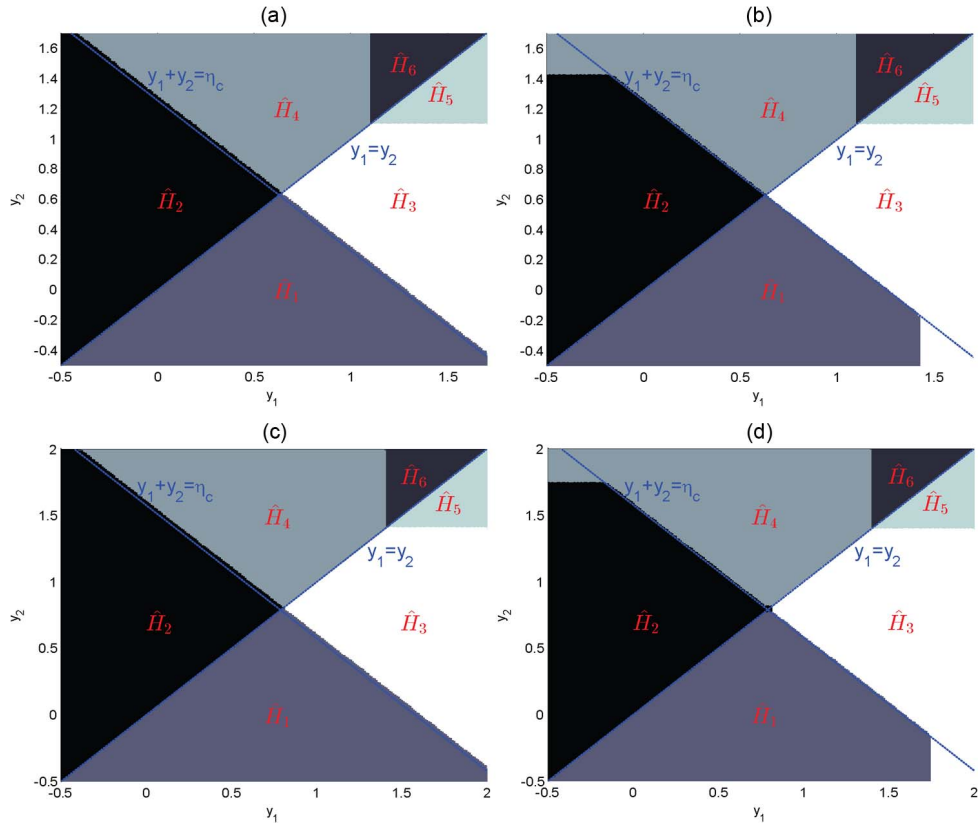

Fig. 5. Decision areas of MAP detection for $\hat{H}$ under various CR and SNR values with $\sigma_x^2 = 0.5$. (a) CR = 8 dB, SNR = 15 dB. (b) CR = 8 dB, SNR = 25 dB. (c) CR = 10 dB, SNR = 15 dB. (d) CR = 10 dB, SNR = 25 dB.

where $\sigma = \sigma_x \sigma_z / \sqrt{\sigma_x^2 + \sigma_z^2}$, and $Q(x) = \int_x^\infty e^{-(x^2/2)} dx$ is the tail probability of the standard normal. Similarly by applying the symmetry property of the hypotheses, the functions $f_{y_{uc},\hat{y}_{uc}|H_2}(y_1, y_2)$, $f_{y_{uc},\hat{y}_{uc}|H_4}(y_1, y_2)$ and $f_{y_{uc},\hat{y}_{uc}|H_6}(y_1, y_2)$ can be readily obtained.

Now that, given the obtained received value $(y_1, y_2)$, the transmission detection in (15) can be calculated by using the derived probability functions in (17), (18), and (22). In Fig. 5, the decision areas of the MAP detection for $\hat{H}$ are plotted under different clipping levels and system SNRs. In this figure, the label $\hat{H}_i (i = 1, 2, \ldots, 6)$ represents the decision area of $\hat{H} = H_i$ given a received pair $(y_1, y_2)$. Once the transmission type has been determined as $\hat{H}$ via the MAP detection, the received subsequently recovers $\mathbf{y}_c$ from the received $\mathbf{y}_{uc}$ as depicted in the diagram of Fig. 2. The recovery procedure is technically an inverse of the RoC procedure in (11) at the transmitter. It simply follows that

$$(y_c[n], \hat{y}_c[n]) = \begin{cases} (y_{uc}[n], 0), & (y_1, y_2) \in \hat{H}_1 \\ (0, \hat{y}_{uc}[n]), & (y_1, y_2) \in \hat{H}_2 \\ (\eta_c + \hat{y}_{uc}[n], 0), & (y_1, y_2) \in \hat{H}_3 \\ (0, \eta_c + y_{uc}[n]), & (y_1, y_2) \in \hat{H}_4 \\ ((1+\alpha)\eta_c, 0), & (y_1, y_2) \in \hat{H}_5 \\ (0, (1+\alpha)\eta_c), & (y_1, y_2) \in \hat{H}_6. \end{cases} \quad (23)$$

Having the recovered data $\mathbf{y}_c$, the receiver applies FFT to complete the OFDM demodulation.

## 4. Simplified Detection Method

Although that the MAP detection is optimal in terms of determining $\hat{H}$, the complexity of the MAP detection is considerable since it needs to calculate the probability function $g(H_i, y_1, y_2)$ which is much involved as shown in the above subsection. For efficient detection in many applications, it is necessary to further present a low complexity detection method for the proposed RoC-ACO-OFDM.

By observing the decision areas in Fig. 5, it shows that the decision areas of most hypotheses are closely related to the two lines $y_1 = y_2$ and $y_1 + y_2 = \eta_c$ regardless of the CR and SNR values. This motivate us to simplify the MAP detection by directly checking the two conditions $y_1 y_2$ and $y_1 + y_2 \eta_c$. Furthermore, it is reasonable to suppose that data recovery under the hypothesis region $\hat{H}_5$ should not affect the performance significantly even if the detected hypothesis drops in $\hat{H}_3$. This is because $(\eta_c + \alpha \eta_c, 0) \approx (\eta_c + \hat{y}[n], 0)$ if one follows the recovery rules in (23) given $(\eta_c, \alpha \eta_c)$ is transmitted. Based on these considerations, we simplify the MAP detection in (15) by merging the six hypotheses into four ones. In this way, the decision areas of the four hypotheses can be directly obtained by only checking a few basic relationships of $y_1$, $y_2$, and $\eta_c$.

Given a pair of received values $(y_{uc}[n], \hat{y}_{uc}[n]) = (y_1, y_2)$, the newly simplified decision areas are defined by

$$\hat{D}_1 : \begin{cases} y_1 > y_2 \\ y_1 + y_2 \leq \eta_c, \end{cases} \quad \hat{D}_2 : \begin{cases} y_1 \leq y_2 \\ y_1 + y_2 < \eta_c \end{cases}$$
$$\hat{D}_3 : \begin{cases} y_1 > y_2 \\ y_1 + y_2 > \eta_c, \end{cases} \quad \hat{D}_4 : \begin{cases} y_1 \leq y_2 \\ y_1 + y_2 > \eta_c. \end{cases} \quad (24)$$

After the decision region of $(y_1, y_2)$ has been determined according to (24), the corresponding recovery procedure for the merged detection regions becomes

$$(y[n], \hat{y}[n]) = \begin{cases} (y_{uc}[n], 0), & (y_1, y_2) \in \hat{D}_1 \\ (0, \hat{y}_{uc}[n]), & (y_1, y_2) \in \hat{D}_2 \\ (\eta_c + \hat{y}_{uc}[n], 0), & (y_1, y_2) \in \hat{D}_3 \\ (0, \eta_c + \hat{y}_{uc}[n]), & (y_1, y_2) \in \hat{D}_4. \end{cases} \quad (25)$$

Fig. 6 plots the merged decision areas for the simplified detection method. It is found that the MAP decision regions $\{\hat{H}_3, \hat{H}_5\}$ and $\{\hat{H}_4, \hat{H}_6\}$ in Fig. 5 are merged as the new decision regions $\hat{D}_3$ and $\hat{D}_4$ in Fig. 6, respectively. The simplified decision regions matches well with the MAP decision regions except for the upper left and lower right corners. By comparing the subfigures in Fig. 5, the difference becomes more pronounced as SNR grows. Fortunately, the difference of decision areas does not lead to a noticeable degradation in terms of the BER performance. Take the upper left corner case, i.e., $(y_1 < 0, y_2 > \eta_c)$ for example. In this case, the received samples are detected with hypothesis $H_4$ according to the MAP detection, and hence the recovered pair equals $(0, \eta_c + y_1)$ according to (23). While with the simplified detection method, the received sample pair locates in the region of $\hat{D}_2$ and it generates the recovered pair as $(0, y_2)$. Consider two cases: 1) If the amplitude of $y_1$ is small and $y_2 \approx \eta_c$, we get $\eta_c + y_1 \approx y_2$; 2) If $y_1$ is negative with a large amplitude or $y_2$ is significantly larger than $\eta_c$, the probability of occurrence is obviously marginal. Therefore, under both cases, the performance may not be significantly affected. Moreover, the near optimal performance of the proposed simplified detection method can also be verified under asymptotically high SNRs. It is not hard to observe that the optimal MAP decision reduces to the simplified one when SNR grows to infinity. In the following section, the near optimal performance is also validated by simulation results. The RoC-ACO-OFDM with simplified detection method achieves almost the same BER performance as the MAP detection does.

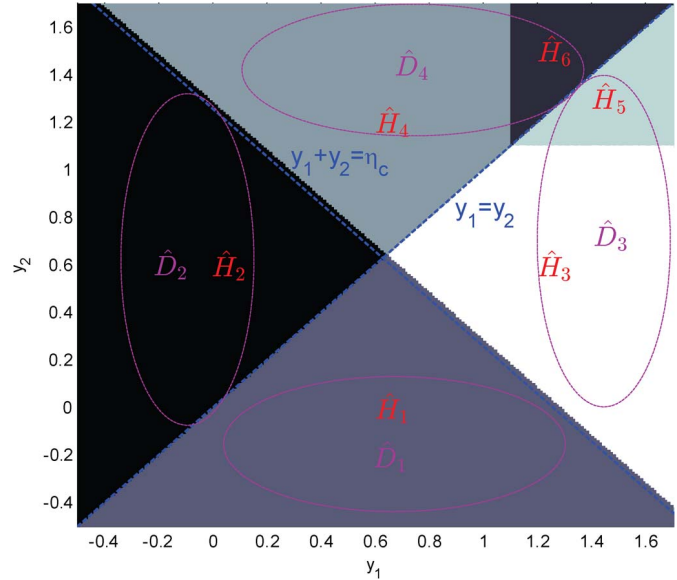

Fig. 6. Decision areas $\hat{D}_i$ of simplified detection.

## 5. Simulation Results

In this section, we present some numerical comparison results between conventional ACO-OFDM and the proposed RoC-ACO-OFDM. Three different ACO-OFDM schemes are tested with PAPR statistics in our simulation and "CR" is utilized to indicate the clipping level. The schemes are listed as follows:

- Ideal ACO-OFDM: The conventional ACO-OFDM implementation without any clipping. When system bit-error rate (BER) is evaluated, no signal distortion is considered with the high PAPR, which theoretically generates an ideal lower bound to BER performance.
- Direct upper clipping: The sample values which are larger than a given clipping level are directly clipped without any other manipulations.
- RoC-ACO-OFDM: The ACO-OFDM samples are upper clipped by applying the proposed RoC procedure.
- Other PAPR reduction methods for optical OFDM: SLM in [15], [16] and precoding based method in [20].

Fig. 7 plots the complementary cumulative distribution function (CCDF) of three ACO-OFDM implementation schemes. From this figure, it shows that for ideal ACO-OFDM one out of every $10^4$ symbols has it PAPR greater than 17.0 dB, but for both direct upper clipping and RoC-ACO-OFDM, one out of every $10^4$ symbols has its PAPR greater than a smaller value of 11.4 dB when the clipping level is selected at CR = 10 dB. This indicates a 5.6 dB PAPR reduction by clipping at CR = 10 dB. By setting different values of CR, the reduction of PAPR can be further controlled as shown in this figure.

It is known that the procedure of clipping inevitably causes performance degradation due to information loss of clipped signals. Fig. 8 compares BER performance of our proposed RoC-ACO-OFDM scheme with existing schemes. The optical channel is assumed to be an LoS channel modeled by the AWGN channel. In this figure, the BER performance of ideal ACO-OFDM serves as a lower bound to the best BER that can be achieved in an ACO-OFDM system because no signal clipping is utilized with the ideal assumption of no distortion due to high PAPR. For the proposed RoC-ACO-OFDM, it significantly outperforms the direct upper clipping approach although both schemes achieve the same PAPR given a fixed CR. Moreover in this figure, we compare the BER performance of RoC-ACO-OFDM by exploiting different detection schemes, i.e., the presented MAP and simplified detection methods respectively in Section 3.2 and Section 4. It can be found that the developed simplified detection methods for RoC-ACO-OFDM achieves almost the same BER performance compared with the complicated MAP detection method, while the implementation of the simplified detection has been shown very efficient with several comparison operations. It is also important to note that, like in any other clipping based technologies, an error floor exists for the RoC-ACO-OFDM. Different tradeoffs are observed between PAPR reduction performance and system BER performance with clipping. A higher CR results in a lower error floor while at the same time it responds to a lower PAPR reduction.

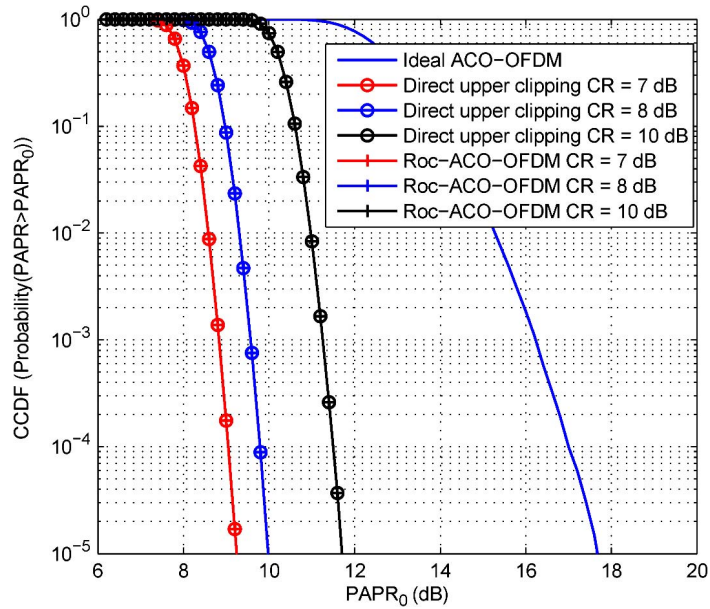

Fig. 7. PAPR CCDF plot under $N = 256$ using 64-QAM.

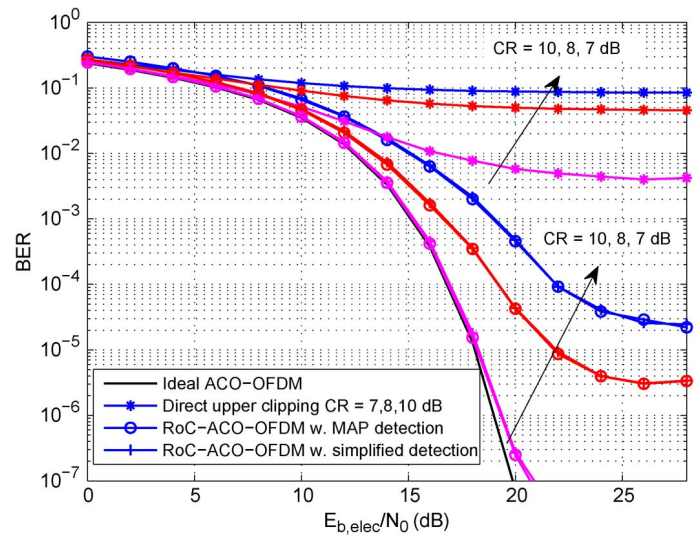

Fig. 8. BER comparisons of different ACO-OFDM schemes under LoS channels with $N = 256$, $CR = \{7, 8, 10\}$ dB, and 64-QAM.

Moreover in Fig. 9, we also examine the BER performance under diffused optical wireless (DOW) channels. The DOW channel is modeled by a sum of a set of positive taps as

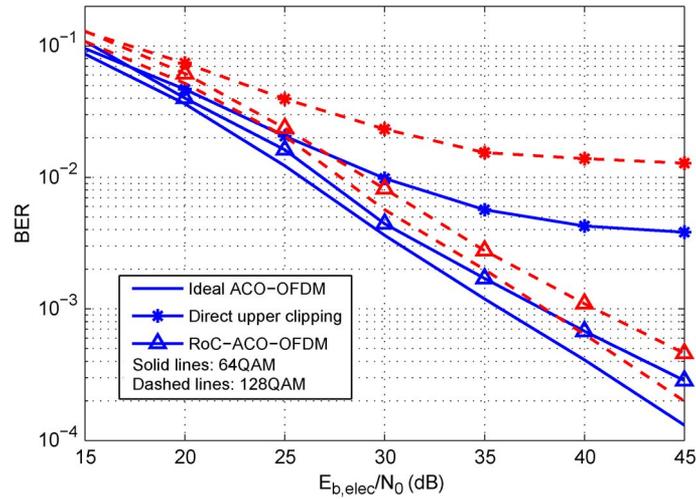

Fig. 9. BER comparisons of different ACO-OFDM schemes under DOW channels with $N = 256$, $CR = 10$ dB, and 64-QAM.

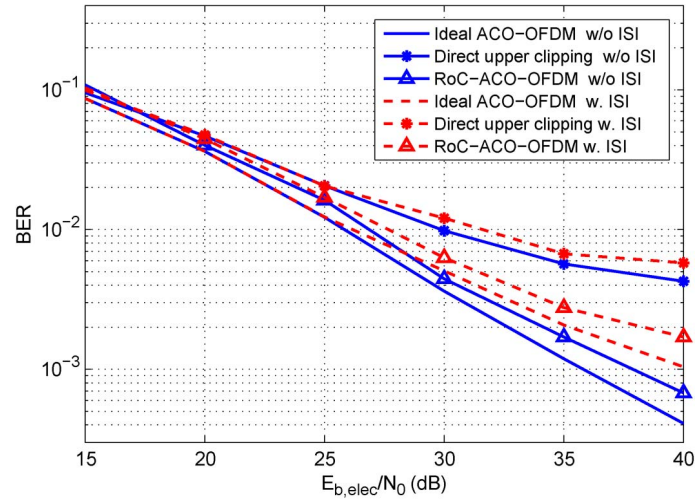

Fig. 10. BER comparison of different ACO-OFDM schemes under DOW channels with/without ISI where $N = 256$, $CR = 10$ dB, and 64-QAM.

follows [27], [28]:

$$h(t) = \sum_{n=0}^{N_t-1} \alpha_n \delta(t - \tau_n) \qquad (26)$$

where $h(t)$ is the channel response at time slot $t$, $\alpha_n > 0$ and $\tau_n$ are the amplitude and time delay of the $n$th path, respectively, and $N_t$ is the number of channel taps. In Fig. 9, we set $N_t = 16$ and the delay is uniformly distributed from 2 to 20 ns. The diffuse fadings follow the exponentially decaying and ceiling bounce models as described in [28], [29] for DOW channels. Similar from the results under LoS channels, Fig. 9 demonstrates a noticeable performance gain by RoC-ACO-OFDM under the diffused channel setup. Note that we tested in Fig. 9 by setting a CP larger than the multipath delay where no inter-symbol interference (ISI) exists. In order to fully evaluate the performance of our proposed RoC-ACO-OFDM, Fig. 10 presents comparison results under the scenario where ISI does exist for the OFDM setup with a shorter CP than the multipath delay. In the simulation, we set a CP of 12-sample long while the maximum delay of the tested multipath channel is 16. From Fig. 10, it is validated that the proposed RoC-ACO-OFDM still outperforms the other schemes although BER performance of all schemes degrades due to the effect of ISI. Therefore, from the comparison results, we can conclude that ISI imposes a similar performance degradation to the proposed scheme as it does to conventional ACO-OFDM.

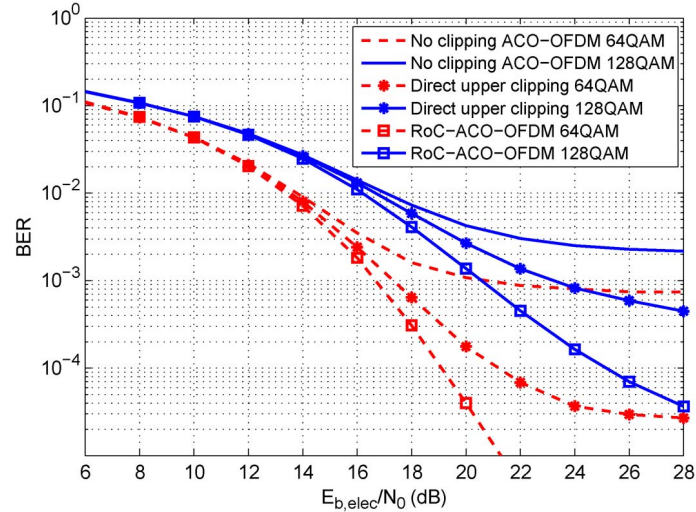

Fig. 11. BER performance comparison under an optical channel including LED nonlinearity.

Since the RoC-ACO-OFDM aims at PAPR reduction, it is also necessary to evaluate the effects of LED nonlinearity. Fig. 11 compares the BER performance of the proposed RoC-ACO-OFDM with two schemes including a naive no clipping scheme and the above-mentioned scheme using direct upper clipping. In this figure, we tested the optical channel by incorporating the LED nonlinearity. We exploit the LED nonlinearity model presented in [22, Eq. (21)]. From this figure, we find that the RoC-ACO-OFDM outperforms the other two schemes because it avoids a high nonlinear distortion by clipping while the information loss due to clipping is minimized. Direct upper clipping achieves better performance than the naive transmission without any clipping. However, the clipping inevitably causes clipped information loss which results in a performance degradation compared to the proposed RoC-ACO-OFDM.

Finally, in Fig. 12, we also compare the PAPR reduction performance of RoC-ACO-OFDM with other PAPR reduction technologies, including the SLM in [15], [16] and precoding schemes in [20]. In an SLM with $R$ phase sequences, any $r$th sequence $\mathbf{u}_r$ has $N/4$ components and $u_{r,n}$ represents the $n$th component of $r$th phase sequence. In our simulation, we choose $u_{r,n} \in \{-1, 1\}$ with equal probability. From this figure, we see that performance of SLM becomes better as $R$ increases at a cost of more computations at the transmitter. Generally, $R$ can not be too large in practical applications. From the comparison results, it shows that RoC-ACO-OFDM achieves a better PAPR reduction performance comparing to SLM with $R$ of median sizes. Precoding is a process wherein the input source symbol sequence $(S(0), S(1), \ldots, S(N/4 - 1))$ is pre-multiplied by a precoding matrix $\mathbf{P}$ [20]

$$\mathbf{P} = \begin{bmatrix} a_{0,0} & a_{0,1} & \cdots & a_{0,N/4-1} \\ a_{1,0} & a_{1,1} & \cdots & a_{1,N/4-1} \\ \vdots & \vdots & \vdots & \vdots \\ a_{N/4-1,0} & a_{N/4-1,1} & \cdots & a_{N/4-1,N/4-1} \end{bmatrix} \tag{27}$$

where $a_{n,k}$ represents an element of the $n$th row and $k$th column and $a_{n,k} = e^{j2\pi nk/N/4}$ for DFT precoding, $a_{n,k} = e^{(j2\pi r/N/4*N/4)((t^2/2)+qt)}$ for ZC precoding, where $t = nN/4 + k$, $r = 3$, and $q = 0$.

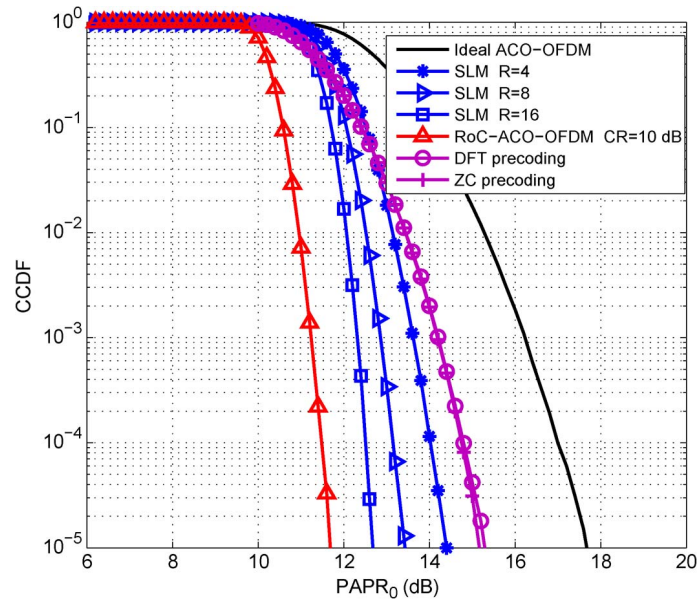

Fig. 12. PAPR performance comparison with SLM using $M = 64$ and $N = 256$ with different iteration times $R = 4, 8, 16$, DFT precoding, and ZC precoding approaches.

By comparing with the precoding scheme using DFT- and ZC-based matrices, we also observe a significant PAPR performance gain by the proposed RoC-ACO-OFDM from the figure results. Note that we choose CR = 10 dB in the simulation where the RoC-ACO-OFDM achieves near ideal BER performance, as revealed in Fig. 8.

## 6. Conclusion

This paper presents a recoverable upper clipping method for ACO-OFDM systems, referred to as RoC-ACO-OFDM. By taking the advantage of special time domain structure of ACO-OFDM symbols, the RoC-ACO-OFDM designs an upper clipping strategy which allows efficient recovery at the receiver side. In this way, the performance degradation due to clipping is minimized and meanwhile the PAPR of the OFDM system is effectively reduced. A significant performance gain is achieved compared with existing clipping schemes. Moreover, both optimal MAP detection and efficient simplified detection methods are developed for RoC-ACO-OFDM. We show that the simplified detection achieves almost the same BER performance as the optical MAP detection does via simulation verifications. This makes the proposed RoC-ACO-OFDM very efficient to implement. Simulation results under both LoS and DOW channels are provided to verify the effectiveness of RoC-ACO-OFDM compared with existing PAPR reduction methods.